\documentstyle[prl,aps,epsf,eqsecnum,floats,preprint,tighten]{revtex}
\begin{document} 
%
\def\pr#1#2#3{ {\sl Phys. Rev.\/} {\bf#1}, #2 (#3)}
\def\prl#1#2#3{ {\sl Phys. Rev. Lett.\/} {\bf#1}, #2 (#3)}
\def\np#1#2#3{ {\sl Nucl. Phys.\/} {\bf B#1}, #2 (#3)}
\def\cmp#1#2#3{ {\sl Comm. Math. Phys.\/} {\bf#1}, #2 (#3)}
\def\pl#1#2#3{ {\sl Phys. Lett.\/} {\bf#1}, #2 (#3)}
\def\apj#1#2#3{ {\sl Ap. J.\/} {\bf#1}, #2 (#3)}
\def\ap#1#2#3{ {\sl Ann. Phy.\/} {\bf#1}, #2 (#3)}
\def\nc#1#2#3{ {\sl Nuo. Cim.\/} {\bf#1}, #2 (#3)}
\newcommand{\beq}{\begin{equation}}
\newcommand{\eeq}{\end{equation}}
\newcommand{\aprime}[1]{#1^\prime}
\newcommand{\daprime}[1]{#1^{\prime\prime}}
\newcommand{\aop}{{A'_0}}
\newcommand{\nop}{{N'_0}}
\newcommand{\da}{\delta a}
\newcommand{\dn}{\delta n}
\newcommand{\dap}{\delta a^\prime}
\newcommand{\dapp}{\delta a^{\prime\prime}}
\newcommand{\dnp}{\delta n^\prime}
\newcommand{\dnpp}{\delta n^{\prime\prime}}
\newcommand{\db}{\delta b}
\newcommand{\dbp}{\delta b^\prime}
\newcommand{\Vdb}{{V(\delta b+\bar b\delta\phi)}}
\newcommand{\ie}{{\it i.e.}}
\newcommand{\eg}{{\it e.g.,}}
\preprint{\vbox{\hbox{UCSD-PTH-00-34} 
     \hbox{IASSNS-HEP-00/88} 
     \hbox{ MIT-CTP-3068}}}

\title{Radion Stabilization by Brane Matter}
\author{Benjam\'\i{}n Grinstein\thanks{e-mail addresses: 
   {\tt bgrinstein@ucsd.edu, nolte@ias.edu, skiba@mit.edu}}, 
Detlef R. Nolte$^\dagger$, and
Witold Skiba$^\ddagger$}
\address{$^*$Department of Physics,  University of California at San Diego,
         La Jolla, CA 92093 \\
         $^\dagger$Institute for Advanced Study, Princeton, NJ 08540 \\
         $^\ddagger$Center for Theoretical Physics, Massachusetts
         Institute of Technology, Cambridge, MA 02139 }
\date{December 2000}

\maketitle
\begin{abstract}
We find a static solution to Einstein's field equations on a five-dimensional
orbifold with a compact $S_1/Z_2$ fifth direction and Poincare
invariant $3+1$ sections. The solution describes a theory with
bulk cosmological constant and 3-branes at the orbifold fixed points
which carry matter density and pressure in addition to tension. The
radius of the fifth dimension is determined by the matter content of the
branes. The ratio of the space and time components of the metric depends 
on the fifth coordinate. Thus, the speed of propagation of massless fields
is path dependent. For example, bulk and brane fields propagate with
different speeds. 
\end{abstract}

\section{Introduction}
\label{intro}
A solution to the hierarchy problem has been proposed in which the
observable universe is a 3-brane at an orbifold fixed point of the
non-factorizable geometry
\beq
\label{eq:RS}
ds^2=e^{-2k|y|}\eta_{\mu\nu}dx^\mu dx^\nu-dy^2.  
\eeq 
The orbifold has fixed points at $y=0$ and $y=y_c$ where there are
branes with tension $V_0=12k$ and $V_c=-12k$, respectively. We will
refer to this as the RS model\cite{Randall:1999ee}. However, the dynamics 
does not determine the value of $y_c$, leaving it a free parameter. 
A solution to this so called ``radion stabilization problem'' has been found 
by adding a bulk scalar field, that is, one that has five-dimensional
dynamics, to the model\cite{Goldberger:1999uk}. Alternative stabilization 
mechanisms have been proposed in \cite{stab-mech}.

In this paper we present a static solution of the field equations when the
branes carry matter density and pressure in addition to tension. The
solution fixes the radius of the fifth dimension. To see how this
works let's first recall why there is a radion stabilization problem
in the RS model. The warp factor in Eq.~(\ref{eq:RS}), $A=-k|y|$, has
a constant slope and, therefore, the jump of the slope across the fixed
point is $-2k$ regardless of where the jump occurs. Now if the warp
factor had a non-constant slope $A'(y)$, then the jump 
$-2A'(y_c)$ does depend on $y_c$. Therefore the radius of the fifth
dimension has to be chosen to accommodate  the jump in energy density
on the brane.

Adding matter to the basic setup of RS is necessary for a description
of cosmology of the model. The cosmology of brane models has been
investigated in a number of papers. A general formulation was given in
Ref.~\cite{Binetruy:2000ut}. The work in 
Refs.~\cite{RScosmo,Csaki:2000mp,Cline:2000tx} 
is concerned
with the cosmology of brane models of the Randall-Sundrum
type\cite{Randall:1999ee}. These papers recognize that the jump in the
warp factors across the branes as implied by Einstein's equations is
enough to give the evolution of the scale factor in a FRW description
of the cosmology on one brane. Therefore there has been little
interest in the behavior of the metric in the bulk, see however 
Ref.~\cite{Cline:2000tx}. It turns out that
investigating bulk solutions can yield interesting results.
The solutions we find exist only for a particular value of
brane separation, which is determined by the matter content of
the branes. Depending on the kind of matter such solutions can 
be either stable or unstable under small perturbations.
The salient feature of our bulk solutions induced by brane matter, 
similar to the solutions in Ref.~\cite{us1}, is that the space and
time components of the metric are not identical as they are in
the RS model without matter. This means that the speed of propagation
of massless fields depends on their trajectory. For example, bulk gravitons
can propagate at a speed different than bulk photons.
Finding out the strength of gravity and masses of bulk and brane fields
in such complicated backgrounds is straightforward using the results of
Ref.~\cite{us2}.

In Refs.~\cite{Binetruy:2000ut,RScosmo,Csaki:2000mp} it is found that 
the scale factor $a$ satisfies
\beq
\label{eq:bdl20}
\left(\frac{{\dot a}}{a}\right)^2 + \left(\frac{\ddot a}{a}\right)=
\frac1{72}V(\rho-3p)-\frac1{36}\rho(\rho+3p),
\eeq
where $V$, $\rho$ and $p$ stand for the tension, matter density and
pressure on the brane respectively. We have set the five-dimensional
gravitational constant to unity. Our static solution is not in
contradiction with this brane equation because the solution demands
that the three parameters $V$, $\rho$ and $p$ are such that the right
hand side vanishes. 

We present our solution in section~\ref{sec:Solution}.  We study
non-static solutions as small perturbations to our solution in
Sec.~\ref{Sec:perturbations} and present our conclusions in
section~\ref{sec:conclusions}.

\section{A Static Solution With Matter}
\label{sec:Solution}
We denote the coordinates of spacetime by $x^A$, $A=0,\dots,4$, and
often use $t=x^0$ and $y=x^4$. The fixed points are at $y=0$ and $y=y_c$. The
class of spherically symmetric metrics we study is parameterized by
three functions of $t$ and $y$ only\cite{Binetruy:2000ut}
\beq
\label{eq:bdlmetric}
ds^2=G_{AB}dx^Adx^B=n^2(t,y)dt^2-a^2(t,y)d\vec x^2 -b^2(t,y)dy^2.
\eeq 
Fixing $y=0$ ($y=y_c$) we see that the metric gives a flat FRW
cosmology on the brane with scale factor $R_0(t')=a(t(t'),0)$
($R_c(t')=a(t(t'),y_c)$) where $dt'=n(t,0)dt$ ($dt'=n(t,y_c)dt$). We
will denote by $g_{\mu\nu}$, with $\mu,\nu=0,\ldots,3$, the induced
metric on the brane.

The action is 
\beq S=\int d^5x\, \sqrt{G}\left[-R-\Lambda \right]
+\int d^4x\,\sqrt{-g}\left[-V_0\right]_{y=0}+ 
\int d^4x\,\sqrt{-g}\left[-V_c\right]_{y=y_c}.  
\eeq 
The constants $\Lambda$, $V_0$ and $V_c$ represent the cosmological
constant in the bulk (5-dimensional space) and on the branes at $y=0$
and $y=y_c$, respectively.  In addition there is matter density and
pressure on the branes, introduced into the field equations directly
by their contribution to the energy momentum tensor:
\beq
T^{AB}=\tilde T^{AB}+\frac{S_0^{AB}}{b}\delta(y)
+\frac{S_c^{AB}}{b}\delta(y-y_c),
\eeq
where $\tilde T^{AB}$ is derived as usual by varying the action with
respect to the metric, and $S^{AB}$ are contributions from perfect
fluids of density $\rho_0$ and $\rho_c$ and pressures $p_0$ and $p_c$
on the branes,
\beq
S^A_B=\mbox{diag}\,(\rho,-p,-p,-p,0).
\eeq

Einstein's equations are
\beq
R^{AB}-\frac12G^{AB}R=\kappa T^{AB}.
\eeq
Here $R^{AB}$ and $R$ are the Ricci tensor and scalar. The
gravitational constant is $\kappa$ and we work in units
of $\kappa=1$. 

For the particular metric (\ref{eq:bdlmetric}) Einstein's equations are
\begin{eqnarray}
3\,\left[ \,\left( 
            (\frac{\dot{a}}{a})^2 + 
            \frac{\dot{a}\,\dot{b}}{a \, b} \right) 
         +  \frac{n^{2}}{b^{2}} \,\left(  -
            \frac{\daprime{a}}{a}        -
            (\frac{\aprime{a}}{a})^2  +
            \frac{\aprime{a}\,\aprime{b}}{a \, b}
            \right) \right] & = &
\frac12n^2\Lambda+ \delta(y)  \frac{n^2}{b} (\frac{1}{2} V_0  + \rho_0)+
          \nonumber\\
& &~~~
          \delta(y-y_c) \frac{n^2}{b} (\frac{1}{2} V_c  + \rho_c)
          \label{Einstein00} \\
3\,\left( 
            \frac{\dot{a}\,\aprime{n}}{a \, n} + 
            \frac{\aprime{a}\,\dot{b}}{a \, b} - 
            \frac{\aprime{\dot{a}}}{a}   \right) & = &
0,  \\
\frac{a^{2}}{n^{2}}\,\left( -
             \frac{\dot{a}^{2}}{a^{2}} -2 
             \frac{\ddot{a}}{a} + 2
             \frac{\dot{a}\,\dot{n}}{a \, n} - 2
             \frac{\dot{a}\,\dot{b}}{a \, b} -
             \frac{\ddot{b}}{b} +
             \frac{\dot{n}\,\dot{b}}{n \, b} \right) &+&
             \frac{a^{2}}{b^{2}}\,\left( 
             \frac{{\aprime{a}}^2}{a^2} +2
             \frac{\daprime{a}}{a} + 2
             \frac{\aprime{a}\,\aprime{n}}{a \, n} - 2
             \frac{\aprime{a}\,\aprime{b}}{a \, b} +
             \frac{\daprime{n}}{n} -
             \frac{\aprime{n}\,\aprime{b}}{n \, b}  \right)
\nonumber   \\  
            =-\frac{a^2}2\Lambda-\delta(y) \,\, \frac{a^2}{b} ( \frac{1}{2} V_0  - p_0)
&-& \delta(y-y_c) \,\, \frac{a^2}{b} ( \frac{1}{2} V_c  - p_c),
  \label{Einstein11} \\
3 \,\left[ %
            \frac{b^{2}}{n^{2}} \,
        \left(  -
             \frac{\dot{a}^{2}}{a^{2}} - 
             \frac{\ddot{a}}{a} + 
             \frac{\dot{a}\,\dot{n}}{a \, n}
         \right) +
          \,\left(
             \left(\frac{\aprime{a}}{a}\right)^2 + 
              \frac{\aprime{a}\,\aprime{n}}{a \, n} 
          \right) 
     \right] & = & -\frac12b^2\Lambda.
  \label{Einstein44} 
\end{eqnarray}       
Here a dot is a shorthand for $\partial/\partial t$ and a prime for
$\partial/\partial y$. The first four equations correspond to the 00,
04, 11 and 44 components of Einstein's equations. Conservation of the
stress-energy tensor gives 
\beq
\label{eq:conserv}
 \dot\rho+3\frac{\dot a}a(\rho+p)=0
\eeq 
on each brane.

We look for static solutions to these equations. One may then
reparametrize the fifth coordinate to enforce $b=1$. The resulting
equations are most easily solved in terms of the warp factors $A$ and
$N$ defined by
\beq
a=\exp(A)\qquad n=\exp(N).
\eeq
For convenience we also introduce $k\equiv\sqrt{-\Lambda/12}$. In the
bulk Einstein equations reduce to
\begin{eqnarray}
A''+2{A'}^2&=&2k^2,\\
2A''+3{A'}^2 +2A'N'+N''+{N'}^2&=&6k^2,\\
{A'}^2 +A'N'&=&2k^2.
\end{eqnarray}
The solution is straightforward. The metric in the bulk is given, for
$y\ge0$, by
\begin{eqnarray}
a^2 &=& a_*^2 \cosh[2k(y-y_*)],\\
n^2 &=& n_*^2 \sinh[2k(y-y_*)] \tanh[2k(y-y_*)],\\
b^2&=&1,
\end{eqnarray}
where $y_*$, $n_*$ and $a_*$ are constants. For related bulk solutions
see Ref.~\cite{Cornell}.

To complete the solution we must examine the field equations at the
brane. The delta functions on the right hand side of
Eqs.~(\ref{Einstein00}) and~(\ref{Einstein11}) must be saturated on
the right hand side by the second derivatives. To this effect we
demand
\begin{eqnarray}
2A'|_{y=0+}&=&-\frac16V_0-\frac13\rho_0,\\
-2A'|_{y=y_c-}&=&-\frac16V_c-\frac13\rho_c,\\
2(2A'+N')|_{y=0+}&=&-\frac12V_0+p_0,\\
-2(2A'+N')|_{y=y_c-}&=&-\frac12V_c+p_c.
\end{eqnarray}
The solution to these are two equations of fine tuning, 
\beq
\label{eq:conds}
144k^2=(V_0+2\rho_0)(V_0-\rho_0-3p_0)=(V_c+2\rho_c)(V_c-\rho_c-3p_c),
\eeq
and two equations fixing the parameters $y_*$ and $y_c$:
\beq
\label{eq:radius}
12k\tanh[2ky_*]=V_0+2\rho_0,\qquad
12k\tanh[2k(y_c-y_*)]=V_c+2\rho_c.
\eeq
The first two conditions,
Eq.~(\ref{eq:conds}) are similar in spirit to the two fine tuning
equations in the RS model that set the square of the tension in each
brane in a fixed proportion to the bulk cosmological constant. 
By comparison, the stabilization mechanism of
Refs.~\cite{Goldberger:1999uk,DFGK} requires only one fine tuning, which
is equivalent to setting the cosmological constant to zero.
The last two equations describe what was alluded to in the introduction,
that in order for the warp factor to jump by the appropriate amount
the location of the brane must be chosen accordingly. Hence the radius
is stabilized, or, should we say, equilibrated (stability is
investigated in the next section).  

Note that positive tension branes ($V>0$) can only exist for $p \leq -\rho$.
If $\frac13 \rho > p > - \frac13 \rho$ on both branes, solutions exist
only for $V_0 < -12 k$ and $V_c < -12 k$. But if $ p > \frac13 \rho$, $V$ 
can be larger or smaller than $- 12 k$.

If we insist, as in the RS model, that $V_0^2=V_c^2=144k^2$ then the
conditions~(\ref{eq:conds}) are identical to the static solution
conditions in Eq.~(\ref{eq:bdl20}). We recover the RS model taking
vanishing matter density and pressure and the
limit $y_*\to\infty$.

However we need not insist on imposing $V_0^2=V_c^2=144k^2$. In fact,
we are now free to chose any tension provided the
conditions~(\ref{eq:conds}) are satisfied and a solution to
Eqs.~(\ref{eq:radius}) with $y_c>0$ can be found.


\section{Small perturbations}
\label{Sec:perturbations}
Armed with the new solutions with static matter density, we proceed to
investigate the time dependence of small matter perturbations. Let us
denote the static solution of the previous section by $n_0=e^N_0$, $a_0=e^A_0$,
and $b_0=1$. We look for solutions to the
field equations,
Eqs.~(\ref{Einstein00})--(\ref{Einstein44}), of the form
\begin{eqnarray}
\label{eq:perturbs}
n & = & n_0 (1+\delta n),\nonumber\\
a & = & a_0 (1+\delta a),\\
b & = & b_0 (1+\delta b).\nonumber
\end{eqnarray}
In addition we set the density on the branes to $\rho_0 +
\delta\rho_0$ and $\rho_c + \delta\rho_c$ and the pressure to $p_0 +
\delta p_0$ and $p_c + \delta p_c$.

We count orders of the perturbative expansion parametrically in
$\delta \rho$ and $\delta p$. That is, we
re-scale $\delta \rho\to\epsilon \delta \rho$, count powers of
$\epsilon$ and set $\epsilon=1$ at the end of the calculation.
In particular this implies
that we make no assumption as to the relative importance of temporal
or spatial derivatives\cite{Csaki:2000mp}. 

To derive the linearized equations in the bulk, we use the
parameterization in Eqs.~(\ref{eq:perturbs}). The 00, 04, 11 and 44
components of Einstein's equations give
\begin{eqnarray}
  \dapp+\aop(4\dap-\dbp)& = & 4k^2\db,\\
  \frac\partial{\partial t}\left((\nop-\aop)\da+\aop\delta b
        -\dap\right) & = & 0,\\
  2\dapp+\dnpp+(6\aop+2\nop)\dap+2(\aop+\nop)\dnp\qquad& &\nonumber\\
     -(2\aop+\nop)\dbp-
\frac1{n_0^2}\left(2\delta\ddot a+\delta \ddot b\right)
      &=&  12k^2\db, \\
  (2\aop+\nop)\dap+\aop\dnp-\frac1{n_0^2}\delta\ddot a
     & = &4k^2\db .
\end{eqnarray}
The solution to these equations gives $\delta b$ and
$\dnp$ in terms of $\delta a$:
\begin{eqnarray}
\label{eq:solbulkb}
\delta b & = &
\frac1{\aop}\left[F+\dap+(\aop-\nop)\da\right],\\
\label{eq:solbulkn}
\dnp & = &
\frac1{\aop}\left[\frac1{n_0^2}\delta\ddot a-(2\aop+\nop)\dap+4k^2\db \right].
\end{eqnarray}
In Eq.~(\ref{eq:solbulkn}) $\db$ is understood as shorthand for the
solution of Eq.~(\ref{eq:solbulkb}). In Eq.~(\ref{eq:solbulkb}) $F$ is
a function of $y$ satisfying
\beq
\label{eq:Fequation}
F'-\frac{A^{\prime\prime}_0-4k^2}{\aop}F=0.
\eeq
It must be observed that $F$ may be discontinuous at $y=0$. In fact,
continuity of $\db$ at $y=0$ requires $F(0+)+F(0-)=0$. The solution to
Eq.~(\ref{eq:Fequation}) is $F(y)=F_*/\sinh(4k(y-y_*))$, with $F_*$ a
constant. 

We connect the bulk solutions for $y>0$ and $y<0$ demanding continuity
of the fields at the brane, $y=0$, and using the jump equations
for the discontinuous derivatives at $y=0$. The
latter give jump conditions for the perturbations
\begin{eqnarray}
\label{eq:jumpa0}
\dap|_{0+} &=&-\frac16\left[(\frac12V_0+\rho_0)\db+\delta\rho_0\right], \\
\label{eq:jumpan0}
(2\dap+\dnp)|_{0+} &= &
\frac12\left[(-\frac12V_0+p_0)\db+\delta p_0\right]. 
\end{eqnarray}
Similarly, the jump equations at the second brane are
\begin{eqnarray}
\label{eq:jumpac}
-\dap|_{y_c-}
&=&-\frac16\left[(\frac12V_c+\rho_c)\db+\delta\rho_c\right], \\
\label{eq:jumpanc}
-(2\dap+\dnp)|_{y_c-} &= &
\frac12\left[(-\frac12V_c+p_c)\db+\delta p_c\right]. 
\end{eqnarray}
In addition, conservation of energy gives, on the brane, 
\beq
\label{eq:energyconperts}
\delta\rho_0+3(\rho_0+p_0)\delta a|_{y=0}=0\qquad{\rm and}\qquad
\delta\rho_c+3(\rho_c+p_c)\delta a|_{y=y_c}=0.
\eeq 
The right hand side of these equations could be a non-zero
constant. However, we set it to zero since we are not interested in
constant shifts in the mass density (since these are accounted for in
the exact solution of Sec.~\ref{sec:Solution}).

The jump equations
at $y=0$, Eqs.~(\ref{eq:jumpa0})--(\ref{eq:jumpan0}), determine
$\dap$ in terms of $\da$ and give an equation for $\da$, namely
\begin{eqnarray}
\frac{1}{n_0^2\nop}\delta \ddot a + \frac{1}{2}\frac\aop\nop\delta p_0
                                  - \frac{1}{6} \delta \rho_0 & = & 0 
\end{eqnarray}
The time dependence of the matter is fixed by the continuity equation,
Eq.~(\ref{eq:energyconperts}). Given an equation of state for the
perturbations, $\delta p_0/\delta\rho_0=w_0$, one can solve this equation.
Let's rewrite the equation as 
\beq	
\label{eq:branediffeq}
\delta \ddot a-\Gamma_0^2 \da=0.
\eeq
Then the coefficient $\Gamma_0^2$ can be expressed in terms of the
initial parameters. In terms of
$t_0\equiv(V_0+2\rho_0)/12k=\tanh(2ky_*)$ we find
\beq
\label{eq:Gammagiven}
\Gamma_0^2=2k^2n^2(y=0)\left[\frac{2}{t_0} - (3 \omega_0 +1) t_0\right]
\left[\frac{1}{t_0} - t_0\right]
\eeq
Hence, $\Gamma_0^2>0$ for  equations of state with
$w_0<1/3$. But for $w_0>1/3$  $\Gamma_0^2$ can be either positive or 
negative depending on the value of $t_0$ with 
$\omega_0 = \frac13 (\frac{2}{t_0^2} -1)$ being the dividing line.

We see that the time dependence of $\da$ on the $y=0$ brane
is exponential for $\omega_0<1/3$, but can be oscillatory for $\omega_0>1/3$:
\beq
\label{eq:da0given}
\da|_{y=0}= c_0e^{\Gamma_0t}+d_0e^{-\Gamma_0t},
\eeq
where $\Gamma_0$ is either real or purely imaginary.
Substituting this solution in the equation for $\dap$ gives
\beq
\label{eq:dap0given}
\dap|_{y=0}=0.
\eeq
This implies that the jump in $F$ vanishes. Therefore $F(y)=0$.

An entirely analogous solution is obtained on the brane at $y=y_c$ by
replacing the subscript ``0'' in
Eqs.~(\ref{eq:branediffeq})--(\ref{eq:da0given}) by the second brane
subscript ``c.''

To extend the solution on the branes into the bulk we must fix the
gauge. 
There is a gauge freedom, that is, reparametrization invariance
consistent with the form of our metric. Starting from the metric
\beq
ds^2=n^2(t',y')dt'^2-a^2(t',y')d\vec x^2 -b^2(t',y')dy'^2,
\eeq
we look for infinitesimal transformations
\begin{eqnarray}
t'&=&t+T(t,y)\\
y'&=&y+Y(t,y)
\end{eqnarray}
that leave the form of the metric invariant and has fixed points at
at $y=0$ and $y=y_c$. Here $T$ and $Y$ are
infinitesimal. The only constraints on these functions come from the
absence of off-diagonal terms in the metric, 
\beq
\label{eq:YTcondition}
n^2T'-b^2 \dot Y=0,
\eeq
and from the fixed points,
\beq
Y(t,0)=Y(t,y_c)=0.
\eeq
Under the gauge transformation the metric variations are
\begin{eqnarray}  
\Delta \dn & = & \nop Y+\dot T ,\\
\Delta \da & = & \aop Y, \\
\label{eq:gaugeTb}
\Delta \db & = & Y'.
\end{eqnarray}
For simplicity we have indicated the variation about a static solution
with $b_0=1$ and $\dot a_0=\dot n_0 =0$.  We would like to use this
gauge freedom to impose the gauge condition
\beq
\label{eq:gauegfix}
\db(y,t)=0.
\eeq
However, this cannot be done in general since
there is only one constant of integration in the gauge condition
(\ref{eq:gaugeTb}). Instead consider covering the space with two different
patches. We label the metric perturbations $\da_1$, $\dn_1$ and $\db_1$
on the patch defined by $0\le y<y_c/2+\epsilon$ and by $\da_2$, $\dn_2$
and $\db_2$ on the patch defined by $y_c/2-\epsilon< y\le y_c$. Given
a solution with metric $\da$, $\dn$ and $\db$ we choose the new
coordinates by choosing
\beq
Y_1=-\int_0^y\db(\hat y,t)d\hat y
\qquad{\rm and}\qquad
Y_2=-\int_{y_c}^y\db(\hat y,t)d\hat y.
\eeq
Thus $\db_1=0$ for $0\le y \le y_c/2+\epsilon$ and $\db_2=0$ for $y\ge
y_c/2-\epsilon$.

The solution for $\da_1$ and $\da_2$ follows immediately from
Eq.~(\ref{eq:solbulkb}) setting $\db=0$:  
\beq
\da_1(y,t)=F_{*1}/4k+\aop \xi_1(t)\qquad{\rm and}\qquad
\da_2(y,t)=F_{*2}/4k+\aop \xi_2(t),
\eeq
where $\xi_{1,2}$ are arbitrary functions of $t$ but independent of
$y$. We can now use our jump conditions to specify these completely:
\beq
\da_1(y,t)=\frac{\aop(y)}{\aop(0+)} \da|_{y=0+}\qquad{\rm and}\qquad
\da_2(y,t)=\frac{\aop(y)}{\aop(y_c-)} \da|_{y=y_c-}.
\eeq

In the overlap region, $|y-y_c/2|<\epsilon$, these
solutions are related by a gauge transformation:
\beq
\da_2-\da_1=\aop\int_0^{y_c}\db(\hat y,t)d\hat y.
\eeq
Thus we obtain
\begin{eqnarray}
\int_0^{y_c}\db(\hat y,t)d\hat y &=&
\frac{\da}{\aop}\Big|_{y=y_c-}-\frac{\da}{\aop}\Big|_{y=0+}\nonumber\\
\label{eq:intbgiven}
 &=& \frac{c_ce^{\Gamma_ct}+d_ce^{-\Gamma_ct}}{k\tanh(2k(y_c-y_*))}
+\frac{c_0e^{\Gamma_0t}+d_0e^{-\Gamma_0t}}{k\tanh(2ky_*)},
\end{eqnarray}
where it is understood that $\Gamma_{0,c}$ can be purely imaginary if
$\omega_{0,c}>\frac13$. 

\section{Discussion and Conclusions}
\label{sec:conclusions}
In Sec.~\ref{sec:Solution} we gave a class of solutions to the field
equations with matter on the two branes. The solutions are static. The
price to pay for time independence is a fine tuning of the amount of
energy on each brane, as expressed in Eq.~(\ref{eq:conds}). These fine
tunings are no worse than the corresponding fine tunings in the RS
model.

The salient feature of the solutions found in Sec.~\ref{sec:Solution}
is that the physical size of the fifth dimension is fixed. This
suggested the exciting possibility that the radion stabilization
problem is not an issue in models with matter on the branes. 

However, static solutions are not cosmologically
acceptable. Nevertheless, one wonders if even for non-static solutions
with matter the radius is stabilized. Short of finding an exact
non-static solution we have exhibited in Sec.~\ref{Sec:perturbations}
an approximate solution by linearizing the field equations around the
static solutions. 

The result of the linearized analysis is summarized by
Eqs.~(\ref{eq:da0given}) and~(\ref{eq:intbgiven}). The first one
establishes that the metric perturbations have exponential time
dependence for equations of state with $\omega=p/\rho<1/3$, but can be
oscillatory for $\omega>1/3$. The second gives the time dependence of
the radius of the space. Indeed, a measure of the radius is
\beq
L=\int\sqrt{-G_{MN}dx^Mdx^N}
\eeq
along a line of constant $t$ and $\vec x$:
\beq
L=y_c+\int_0^{y_c}\db(\hat y,t)d\hat y.
\eeq
For $\omega<1/3$ the radius grows exponentially, at least while the
exponential is small enough that the linearized solution remains a
good approximation. On the other hand, for $\omega>1/3$ the radius
can oscillate about the equilibrium value $y_c$ or can grow exponentially 
depending on the value of $t_0 = \frac{V_0 + 2 \rho_0}{12 k}$.

One gains some understanding of the behavior of the scale factor $a$
on the brane by considering the full, non-perturbative description of
its evolution at an orbifold fixed point. Since we do not wish to
insist that $V^2=144k^2$ we modify Eq.~(\ref{eq:bdl20}) to allow for
an unconstrained tension,
\beq
\label{eq:bdl20modified}
\left(\frac{{\dot a}^2}{a^2}\right) + \left(\frac{\ddot a}{a}\right)=
\frac1{72}V(\rho-3p)-\frac1{36}\rho(\rho+3p) +\frac1{72}V^2 -2k^2,
\eeq
Let the equation of state be $p=w\rho$. Using the equation of
conservation, Eq.~(\ref{eq:conserv}), one has
$\rho=\rho_0a_0^{3(1+w)}/a^{3(1+w)}$, where $\rho_0$ and $a_0$ are the
density and scale factor at a fixed time. Then one can rewrite the
equation for the form factor as the equation for a particle with
displacement $r=a^2$ in a potential, $\ddot r =-U'(r)$, with
\beq
U(r)=C_1 r^{2-\frac32(1+w)}+ C_2 r^{2-3(1+w)} + C_3 r^2,
\eeq
where 
\begin{eqnarray}
C_1&=&-\frac{\rho_0a_0^{3(1+w)}V}{18}\\
C_2&=&- \frac{\rho_0^2a_0^{6(1+w)}}{18}\\
C_3&=&2 k^2-\frac{V^2}{72}
\end{eqnarray}
Recall that for $\omega > -1$ our solutions must have negative tension branes
$V<0$, so $C_1>0$ and $C_2<0$. For $\frac13 > \omega > -\frac13$, $V<-12k$ 
and, therefore, $C_3$ is negative and the only extremum of $U(r)$ is 
a maximum. But for $w>\frac13$, V can be larger or smaller than $-12k$ 
and, therefore, the sign of $C_3$ can be either positive or negative. 
The extremum of $U(r)$ can, therefore, be a maximum or minimum, 
in accordance with our analysis of small perturbations. 

The physical interpretation of the solutions discussed here is not
straightforward. One cannot, as in the case of the RS model, simply
renormalize fields on either brane to cast their kinetic energy term
in standard form, since $n\ne a$. Moreover, a null trajectory parallel
to the branes has $dx/dt=n/a$. Since $n/a=(n_*/a_*)\tanh[2k(y-y_*)]$  
differs between the two fixed points, one can arrange for superluminal
signal travel. For example, on one brane one can send a graviton across
to the other brane, relay the signal along the brane by photons, and
then relay the signal back to the first brane via
gravitons. Neglecting the time of travel between branes, we see that
the interval between emission and reception can be shorter than the
time for a photon to travel between the same two points directly on
the first brane\cite{Chung:2000xg}. A complete discussion is the
subject of a forthcoming paper. 

Since for standard fluids in equilibrium the equation of state must
have $\omega<1/3$, the exponential growth of the scale factor and
radius make this model an unlikely candidate for cosmology. However,
it may be possible to stabilize the radius by modifying the model,
say, by adding matter in the bulk, {\it e.g.}, scalar
fields\cite{Goldberger:1999uk}. The cosmology of such a model may
indeed be acceptable\cite{Csaki:2000mp}.

\bigskip

{\it Note Added}
While this article was being completed three related works have been submitted
to the archives. Ref.~\cite{CEG} discusses general solutions with different
space and time components of the 5-d metric. We have not attempted to
check if an explicit coordinate transformation can convert our bulk
solution to a form used in Ref.~\cite{CEG}. The authors also discuss 
Lorentz symmetry violations due to different propagation speeds of bulk
and brane fields. Ref.~\cite{LutySundrum} uses supersymmetric gauge dynamics
for stabilizing the radius and Ref.~\cite{HofKanPos} uses the Casmir force 
due to a bulk scalar field.

\bigskip

{\it Acknowledgments} 
The work of B.G.\ is supported by the U.S. Department of
Energy under contract No.\ DOE-FG03-97ER40546, the work of D.N.\ 
under contract No.\ DE-FG02-90ER40542, and the work of W.S. under cooperative
research agreement DE-FC02-94ER40818.



\begin{references}

\bibitem{Randall:1999ee}
L.~Randall and R.~Sundrum,
Phys.\ Rev.\ Lett.\  {\bf 83}, 3370 (1999)
[hep-ph/9905221].

\bibitem{Goldberger:1999uk}
W.~D.~Goldberger and M.~B.~Wise,
Phys.\ Rev.\ Lett.\  {\bf 83}, 4922 (1999)
[hep-ph/9907447].

\bibitem{stab-mech}
L. ~Mersini,
``Radion potential and brane dynamics,''
hep-ph/0001017;
H. B.~Kim,
Phys.\ Lett.\  {\bf B478}, 285 (2000)
[hep-th/0001209];
J. ~Garriga O. ~Pujolas and T. ~Tanaka,
``Radion effective potential in the brane-world,''
hep-th/0004109;
U.~Gunther and A.~Zhuk,
``A note on dynamical stabilization of internal spaces in 
multidimensional cosmology,''
hep-ph/0006283;
W.~D.~Goldberger and I.~Z.~Rothstein,
Phys.\ Lett.\ {\bf B491}, 339 (2000)
[hep-th/0007065];
I.~Brevik, K.~A.~Milton, S.~Nojiri and S.~D.~Odintsov,
``Quantum (in)stability of a brane-world AdS(5) universe at nonzero  temperature,''
hep-th/0010205.

\bibitem{Binetruy:2000ut}
P.~Binetruy, C.~Deffayet and D.~Langlois,
Nucl.\ Phys.\  {\bf B565}, 269 (2000)
[hep-th/9905012].

\bibitem{RScosmo}
C. ~Csaki, M. ~Graesser, C. ~Kolda and J. ~Terning,
Phys.\ Lett.\  {\bf B462}, 34 (1999)
[hep-ph/9906513];
J.~M.~Cline, C.~Grojean and G.~Servant,
Phys.\ Rev.\ Lett.\ {\bf 83}, 4245 (1999)
[hep-ph/9906523];
D.~J.~Chung and K.~Freese,
Phys.\ Rev.\ D {\bf 61}, 023511 (2000)
[hep-ph/9906542];
P.~Binetruy, C.~Deffayet, U.~Ellwanger, and D.~Langlois,
Phys.\ Lett.\  {\bf B477}, 285 (2000)
[hep-th/9910219];
E.~E.~Flanagan, S.~H.~Tye and I.~Wasserman,
Phys.\ Rev.\  {\bf D62}, 044039 (2000)
[hep-ph/9910498];
H.~Stoica, S.~H.~Tye, and I.~Wasserman, Phys.\ Lett.\  {\bf B482}, 205 (2000)
[hep-th/0004126];
R.~N.~Mohapatra, A.~Perez-Lorenzana and C.~A.~de Sousa Pires,
``Cosmology of brane-bulk models in five dimensions,''
hep-ph/0003328.


\bibitem{Csaki:2000mp}
C.~Csaki, M.~Graesser, L.~Randall and J.~Terning,
Phys.\ Rev.\  {\bf D62}, 045015 (2000)
[hep-ph/9911406];

\bibitem{Cline:2000tx}
J.~M.~Cline and H.~Firouzjahi,
Phys.\ Lett.\ {\bf B495}, 271 (2000)
[hep-th/0008185].

\bibitem{us1} 
B.~Grinstein, D.~Nolte and W.~Skiba,
Phys.\ Rev.\  {\bf D62}, 086006 (2000)     
[hep-th/0005001].

\bibitem{us2}
B.~Grinstein, D.~Nolte and W.~Skiba,
``On a covariant determination of mass scales in warped backgrounds,''
hep-th/0012074.

\bibitem{Cornell}
S. H. H.~Tye and I.~Wasserman,
``A brane world solution to the cosmological constant problem,''
hep-th/0006068;
E.~Flanagan, N.~Jones, H.~Stoica, S. H. H.~Tye and I.~Wasserman,
``A Brane World Perspective on the Cosmological Constant and the 
Hierarchy Problems,''
hep-th/0012129.

\bibitem{DFGK}
O.~DeWolfe, D.~Z.~Freedman, S.~S.~Gubser and A.~Karch,
Phys.\ Rev.\  {\bf D62} (2000) 046008
[hep-th/9909134].

\bibitem{Chung:2000xg}
D.~J.~Chung and K.~Freese,
Phys.\ Rev.\ D {\bf 62}, 063513 (2000)
[hep-ph/9910235].

\bibitem{CEG}
C.~Csaki, J.~Erlich and C.~Grojean,
``Gravitational Lorentz Violations and Adjustment of the Cosmological 
Constant in Asymmetrically Warped Spacetimes,''
hep-th/0012143.

\bibitem{LutySundrum}
M.~Luty and R.~Sundrum,
``Hierarchy Stabilization in Warped Supersymmetry,''
hep-th/0012158.

\bibitem{HofKanPos}
R.~Hofmann, P.~Kanti and M.~Pospelov,
``(De-)Stabilization of an extra dimension due to a Casimir force,''
hep-ph/0012213.

\end{references}
\end{document}